\documentclass{article}

\usepackage{PRIMEarxiv}

\usepackage[utf8]{inputenc} 
\usepackage[T1]{fontenc}    
\usepackage{hyperref}       
\usepackage{url}            
\usepackage{booktabs}       
\usepackage{amsfonts}   
\usepackage{amssymb}
\usepackage{nicefrac}       
\usepackage{microtype}      
\usepackage{lipsum}
\usepackage{fancyhdr}       
\usepackage{graphicx}   
\usepackage{capt-of}         
\graphicspath{{media/}}     

\usepackage{lscape}
\usepackage{titlesec}
\usepackage{adjustbox}
\usepackage{array}
\usepackage{tabularx}
\usepackage{authblk}

\titleformat{\paragraph}
{\normalfont\normalsize\bfseries}{\theparagraph}{1em}{}
\titlespacing*{\paragraph}
{0pt}{3.25ex plus 1ex minus .2ex}{1.5ex plus .2ex}

\title{A GRAPHICAL APPROACH FOR BRAIN HAEMORRHAGE SEGMENTATION}

\author[3]{Dr. Ninad Mehendale }
\author[3]{Pragya Gupta }
\author[1]{Nishant Rajadhyaksha}
\author[1]{Ansh Dagha }
\author[1]{Mihir Hundiwala}
\author[1]{Aditi Paretkar}
\author[3]{Sakshi Chavan}
\author[2]{Tanmay Mishra}
\affil[1]{Department of Computer Science and Engineering, K.J. Somaiya College of Engineering}
\affil[2]{Department of Mechanical Engineering, K.J. Somaiya College of Engineering}
\affil[3]{Department of  Electronics and Telecommunication Engineering, K.J. Somaiya College of Engineering}

\begin{document}
\maketitle

\begin{abstract}
Haemorrhaging of the brain is the leading cause of death in people between the ages of 15 and 24 and the third leading cause of death in people older than that. Computed tomography (CT) is an imaging modality used to diagnose neurological emergencies, including stroke and traumatic brain injury. Recent advances in deep learning and image processing have utilised different modalities like CT scans to help automate the detection and segmentation of brain haemorrhage occurrences. In this paper, we propose a novel implementation of an architecture consisting of traditional Convolutional Neural Networks(CNN) along with Graph Neural Networks(GNN) to produce a holistic model for the task of brain haemorrhage segmentation. GNN's work on the principle of neighbourhood aggregation thus providing a reliable estimate of global structures present in images. GNN's work with few layers thus in turn requiring fewer parameters to work with. We were able to achieve a dice coefficient score of around 0.81 with limited data with our implementation.
\end{abstract}


\section{Introduction}
The brain is one of the most vital components of a functioning neural system for humans. Brain haemorrhages are a life-threatening condition and can be caused by either physical injury or medical conditions such as high blood pressure \cite{doi:10.1056/NEJMoa1603460} or an aneurysm. The mortality rate for brain haemorrhages is around 49.5\% \cite{2}. Haemorrhages can be categorized into five types depending on their location and shape:  epidural, intraventricular, subarachnoid, intraparenchymal, subdural. The diagnosing of haemorrhages before treatment is a crucial step.
Practitioners often spend a lot of time going through the medical images to segment areas with haemorrhages. It is of utmost importance to automate the segmentation of haemorrhages so that experts can start the treatment of patients earnestly. C.T scans are often used for segmenting haemorrhages as they are cheaper than Magnetic Resonance Imaging (M.R.I) images and are easier to acquire. The CT procedure combines information from multiple X-ray images taken at different angles to produce cross-sectional slices. Attenuation, which is measured in Hounsfield Units (HU) and ranges from -1000 for air to 0 for water and +2000 for dense bones, indicates how much the X-ray was attenuated. The type of tissue at each point can be determined from the intensity value of voxels. Various brain tissues are shown in the grayscale image by selecting different window parameters e.g., stroke window, bone window \cite{3}. Using the brain window, the CT scan images show that ICH areas appear hyperdense with relatively ill-defined structures. Convolutional neural networks (CNN) are excellent at automating multiple classifying and segmenting tasks in images due to recent advances \cite{4}. Identifying brain lesions from non-contrast head computed tomography (CT) images can prove particularly challenging due to the multiparametric heterogeneity found in the images. Therefore we developed a novel model to help assist in the automation of Segmenting Brain haemorrhages on CT scans. Our network consists of a CNN module, A GNN module and finally an inference module. The GNN module is a novel addition as no other paper has implemented GNN's for the task of brain haemorrhage segmentation to date as to our knowledge. We have chosen to utilise GNN's as a key component in our model as it helps reduce the parameter count for the model making it suitable for the size of the data set we train our model on and it helps detect global structures which the CNN model alone cannot \cite{zhou2021graph}. The only publicly available data set containing segmentation masks for haemorrhages is the one cited in the paper by M.Hassayeni et al. \cite{6}. We further describe different loss functions used commonly for the task of semantic segmentation on brain haemorrhages.

\section{Literature Review}
\label{sec:headings}
Semantic Segmentation refers to the task of labelling each pixel of an image with a corresponding class of what is being represented. U-Net \cite{ronneberger2015unet} is the standard architecture used for the process of semantic segmentation. Semantic segmentation consists of a contracting path followed by an expansive path thus giving it the u-shape. As the contracting path shrinks, the spatial information is reduced while the feature information increases. The expansive pathway combines the feature and spatial information by concatenating the high-resolution features from the contracting path. Deep Learning has made significant headway in the field of medicine. Mohammed Havei et al. presented a novel architecture for automatic brain tumour segmentation \cite{HAVAEI201718}. The authors developed a cascading architecture in which the output of a basic CNN is used as an additional source of information for subsequent CNN layers. Results reported on the 2013 BRATS test data-set reveal that their architecture improves over the established models while being over 30 times faster. The work of Kritannawong et al. focused on predictive outcomes for the problems of classification of coronary artery disease, heart failure, strokes and cardiac arrhythmia \cite{8}. The authors utilised several machine learning algorithms to test for several cardiovascular diseases. The paper described a pooled output AUC score of about 0.91 for the classification of cardiovascular diseases. Considerable work has been done to help solve problems relating to brain haemorrhages. Models are generally developed for tasks relating to segmentation and classification. The majority of classification works have focused on binary detection tasks. Arbabshirani et al. Presented a CNN based architecture to detect the presence of intracranial haemorrhages in a CT scan image which was trained on 37,074 studies and reported an AUC under the ROC curve of 0.846 \ to cite{9}. A multi-class classifier was developed by Chang et al. and achieved an accuracy of 97\% for the detection of epidural/subdural, subarachnoid and intraparenchymal haemorrhages on non-contrast CT \cite{pmid30049723}. Brain haemorrhage segmentation is a well-known problem in deep learning literature. Yuh and Colleagues used traditional machine learning methods to detect intra-cranial haemorrhages based on their location shape and volume \cite{7}. In their study, the authors optimized the threshold value for 210 CT scans of subjects with suspected TBI using retrospective samples of 33 CT scans. Their algorithm achieved results of about 59\% specificity and 98\% sensitivity. A model following the principle of weighted greyscale histogram hierarchical classification was trained on CT scan images and obtained an average accuracy of 95\% for segmenting different types of haemorrhages \cite{SHAHANGIAN2016217}. Hassayeni et al. reported a dice score of 0.31 for segmentation tasks performed using a U-Net based architecture \cite{hssayeni2019intracranial}. They collected a dataset of 82 CT Scans containing segmentation masks marked by an experienced radiologist \cite{6}. Jadon et al. described a 2D U-Net++ architecture that achieved a dice score of 0.94 for intraparenchymal haemorrhage cases \cite{2020}. They further reported the 0.05 increase in the reported dice score after optimizing their model using focal-Tversky loss \cite{abraham2018novel}. Lin et al. proposed a self-supervised learning method where they used a Noisy Student learning technique in conjunction with a supervised fully convolutional neural network named PatchFCN \cite{lin2021noisy}. The authors reported an average pixel accuracy of around 84\% on the CQ500 dataset \ to cite{chilamkurthy2018development}. An unsupervised approach was presented in which mixture models were trained on CT scan images \cite{19}. The authors stated that their algorithm utilizes the fact that the properties of haemorrhage and healthy tissues follow different distributions, and therefore an appropriate formulation of these distributions allows the separation of haemorrhages through an Expectation-Maximization process. Graph Neural Networks have recently been used in medical imaging cases. Juarez et al. presented an architecture consisting of GNN in conjunction with CNN for the task of chest airways segmentation using CT scans \cite{Juarez2019AJ3}. As to our knowledge, no study has implemented a model incorporating GNN's for the task of brain haemorrhage segmentation. 

\section{Methodology}

In the following sections, we provide a detailed description of the methodology followed during the study.

\subsection{Dataset}
\begin{figure}[t]

\includegraphics[width=8cm]{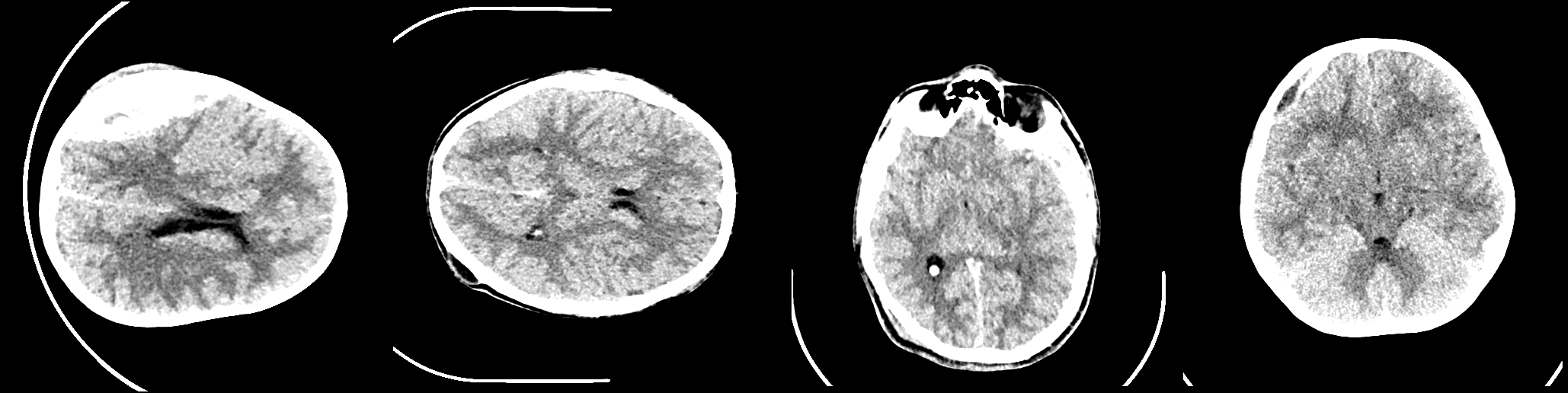}
\centering
\caption{Images of Brain haemorrhage CT-Scan Slices}
\label{fig:Images of CT Scans}
\end{figure}

\begin{figure}[t]

\includegraphics[width=8cm]{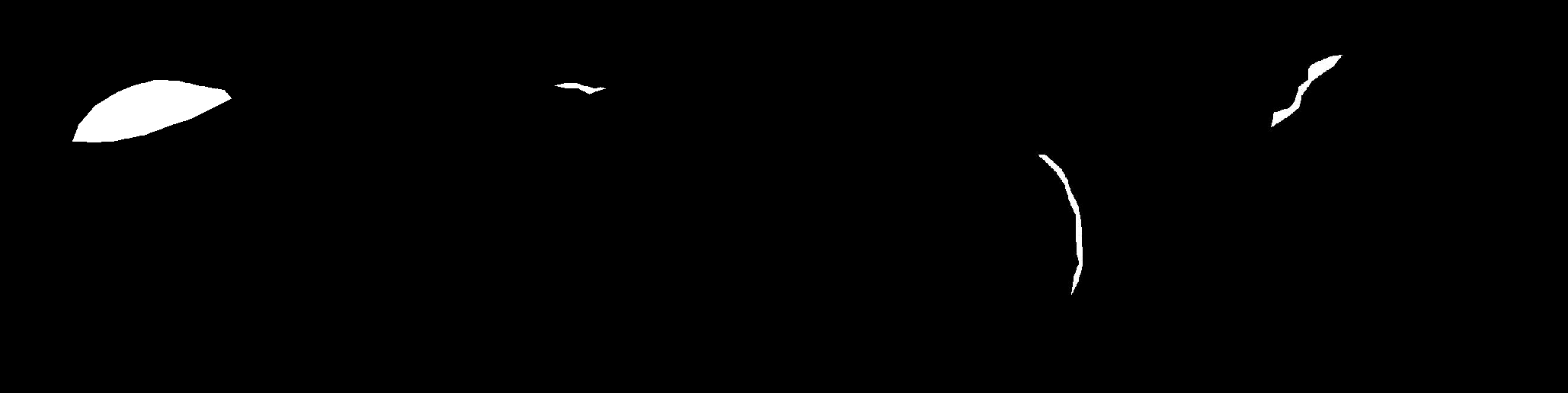}
\centering
\caption{Images of Brain haemorrhage CT-Scan Segmentation Masks}
\label{fig:ImagesCTMAsks}
\end{figure}

A publicly available dataset \cite{6} containing 82 CT scans were utilised for this study. The dataset includes 36 scans of patients diagnosed with intracranial haemorrhages of types Subarachnoid, Epidural, Intraventricular, Subdural and  Intraparenchymal. Approximately 30 slices of 5 mm thickness are used for each CT scan. The mean and standard deviation of the age of the patients were 27.8 and 19.5, respectively. 46 of the patients were male, and 36 were female. Two senior radiologists segmented each slice of all non-contrast CT scans for different brain haemorrhages. The radiologists had no access to the historical clinical records of patients. Intra-cranial regions were delineated in each CT slice by the radiologists. Siemens Medical Solutions syngo was first used to process the CT DICOM files. Two videos were saved (avi format) using bone and brain windows. Gray Scale 650X650 images (jpg format) were finally produced for the brain window. Examples of CT scan slices and corresponding segmentation masks are shown in figures \ref{fig:Images of CT Scans},\ref{fig:ImagesCTMAsks} respectively.

\begin{figure}[h]

\includegraphics{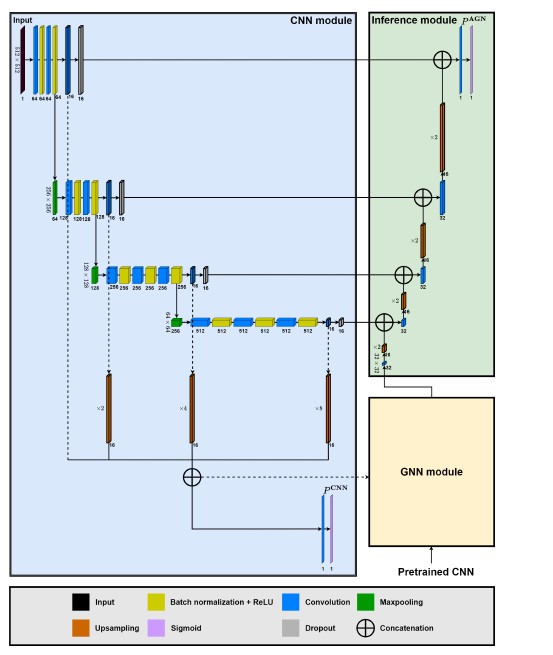}
\centering
\caption{Model Architecture}
\label{fig:modelarch}
\end{figure}

\subsection{Model Architecture}
In this section, we describe the architecture of our model. Our architecture consists of three modules namely the CNN module, the GNN module and the inference module. The architecture is shown in figure \ref{fig:modelarch}.

\subsubsection{Convolution Neural Network Module}

The general principle followed by the CNN module is that of the contracting path of U-Net. The model architecture is also heavily inspired by the Deep Retinal Image Understanding network (DRIU) \cite{2016} which is based on the model built by Visual Geometry Group(VGG). The CNN module tries to encode the features of a 512X512 grey-scale image to a lower-dimensional representation whilst finally producing a single channelled probability map which results in approximated segmentation results. There are 14 convolution layers in total in the CNN module. Each convolutional layer is followed by a sequence of an activation and batch normalisation layer. The batch normalisation layer provides several benefits. Without a batch normalisation layer, there is an internal covariate shift between layers which forces a change in the distribution of features in the output. Further dropout layers are used before skip connections to help regularise the model and keep a check on over-fitting. There are 3 max-pooling layers in total in the CNN module. To extract sharper features in the tensors, we reduce their size during the max-pooling process. after the sequences of convolutional layers batch normalisation layers activation layers and max-pool layers the outputs are concatenated to with up-sampling layers and residual connections. The residual connections help in the smoothing of the loss landscape which helps the model to generalise better \cite{li2018visualizing}. the concatenated outputs are then passed through a convolutional layer followed activation function to produce a 512X512 sized single channelled probability map. Our model incorporates rectified linear activation unit (ReLU) as the default activation function throughout the CNN module except for the final concatenated output which is activated by the sigmoid activation function.

\subsubsection{Graph Neural Network Module}

Here we begin by providing some background to describe GNN's.
Some common notations used to describe graphs are listed in table \ref{tab:NotationTable}.

\begin{table}[h]

\includegraphics{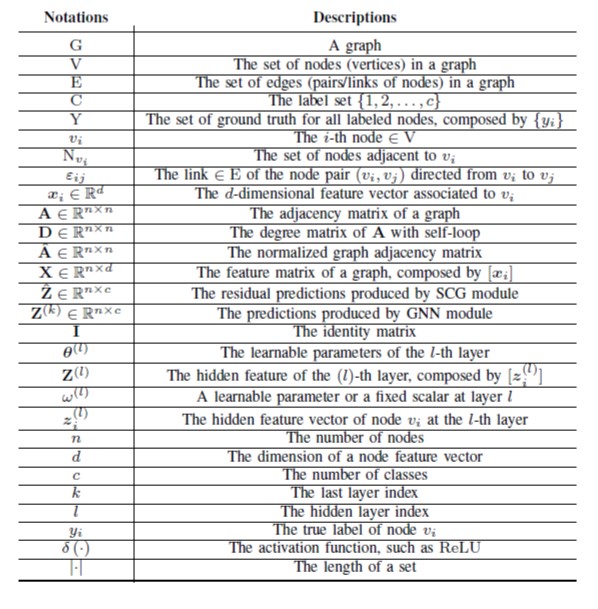}
\centering
\caption{Graph Notations}
\label{tab:NotationTable}
\end{table}

We propose a network structure consisting of Self Constructing Graph (SCG) Network \cite{liu2020selfconstructing} in conjunction with a GNN layer namely Graph Convolutional Network layer (GCN) \cite{kipf2017semisupervised}.

\paragraph{SCG Network}

\begin{figure}[h]

\includegraphics{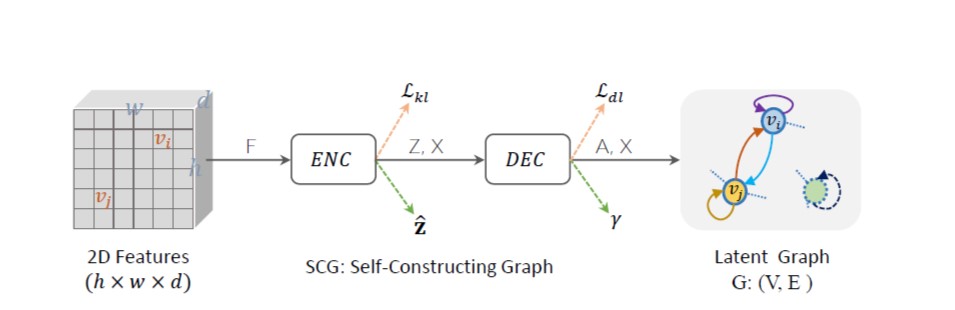}
\centering
\caption{Self Constructing Graph Network}
\label{fig:SGC}
\end{figure}

The SGC network aims to learn latent representations of graphs by capturing global context information from 2D feature maps directly without relying on prior knowledge. The SCG model is represented by  the figure \ref{fig:SGC}. 

The SCG network draws inspiration from variational graph auto-encoders \cite{kipf2016variational}. The SCG network contains an Encoder module and a Decoder module. The input F  \begin{math} \in \mathbb{R}^{h\times w\times d} \end{math} is encoded to a latent embedding space  Z \begin{math}  \in \mathbb{R}^{n\times c} \end{math} by the encoder module where \emph{h,w,d,n,c} represent height, width, channels in the input image, number of nodes and node features respectively. Graph representations G: (V,E) \begin{math} \iff \end{math} (A,X) are generated by the decoder module by decoding the latent embeddings Z. Further to aid the training process a Kullback-Leibler divergence (\begin{math} \mathcal{L}_{kl} \end{math}) and diagonal log penalty (\begin{math} \mathcal{L}_{dl} \end{math}) terms are introduced as a regularisation measure along with adaptive enhancement measure \begin{math} \gamma \end{math}. In essence the SCG network presents graph embeddings G for a corresponding image which can be represented by equation .
\begin{equation} SCG\left( F\right) =\left( G,\mathcal{L}_{kl}, \mathcal{L}_{dl},\gamma \right) \end{equation}

The encoder module first reduces the spatial dimensions of input image F to $F^{'}$ \begin{math} \in \mathbb{R}^{\overline{h}\times \overline{w}\times d} \end{math} by utilising parameter-free pooling methods (e.g. adaptive average pooling). $F^{'}$ is then transformed to obtain X which is a 2D feature matrix containing n nodes  where n = \begin{math} \overline{h} \times \overline{w} \end{math}. The encoder module learns a mean matrix M \begin{math} \in \mathbb{R}^{n \times c} \end{math} and a standard deviation matrix \begin{math} \Sigma \in \mathbb{R}^{n \times c} \end{math}  by using convolutional layers.

\begin{equation}
    M=Conv_{3\times 3}\left( {F^{'}}\right) 
\end{equation}

\begin{equation}
    \log  (\Sigma) =Conv_{1\times 1}\left( {F^{'}}\right) 
\end{equation}

We output the \begin{math} \log(\Sigma) \end{math} to ensure stability during the training process. The encoder module then performs a reparameterization operation to obtain latent embeddings Z representing the image.

\begin{equation}
    Z= M + \Sigma \cdot \Upsilon
    \label{eqn:repara}
\end{equation}

\begin{math} \Upsilon \end{math} in equation \ref{eqn:repara} is a noise variable utilised during training and is sampled from a standard normal distribution. An isotropic centered Gaussian multivariate prior distribution is used to regularise latent variables, by minimizing the Kullback-Leibler divergence loss.

\begin{equation}
    \mathcal{L}_{kl}=-\frac{1}{2nc} \cdot \sum_{i=1}^{n}  \sum_{j=1}^{c}(1 + \log(\Sigma_{ij})^{2} - M_{ij}^{2} - (\Sigma_{ij})^{2})
\end{equation}

The decoder block generates the graph adjacency matrix A by taking an inner product between latent embeddings as A = ReLU \begin{math}(Z \cdot Z^{T}\end{math}). A represents weighted edges hence A \begin{math}_{ij} = \end{math} A\begin{math}_{ji} > 0 \end{math} indicates the presence of an edge. The decoder module builds the graph from information received by connecting similar node representations together. It does so by measuring similarity between node features such that similar neighbourhood regions exchange information.
We introduce another regularisation term named diagonal log regularisation defined in equation \ref{eqn:dl}.

\begin{equation}
    \mathcal{L}_{dl} = - \frac{\Upsilon}{n^{2}}\cdot \sum_{i=1}^{n} \log(|A_{ii}|_{[0,1]}+\epsilon)
    \label{eqn:dl}
\end{equation}

The index \begin{math}_{[0,1]}\end{math} represents the process of clamping A\begin{math}_{ii}\end{math} to [0,1], \begin{math} \epsilon \end{math} is a small positive infinitesimal scalar (e.g, $\epsilon$ = $10^{-7}$) and $\Upsilon$ is the adaptive factor computed as :

\begin{equation}
    \Upsilon=\sqrt{1+\frac{n}{\sum ^{n}_{i=1}A_{ii}+\varepsilon }}
\end{equation}

To stabilise training we introduce the adaptive enhancement factor to the adjacency matrix which is indicated in equation \ref{eqn:adjmat}.

\begin{equation}
    A = (A + \Upsilon \cdot diag(A))
    \label{eqn:adjmat}
\end{equation}

\paragraph{GCN Layer}
The general idea of GCN is to apply convolution over a graph. Convolution on a graph is defined on its neighbouring nodes. The principle of of GCN was that based on first order approximations of spectral GNN's \cite{HAMMOND2011129}. A GCN layer performs linear transformation over a one-hop neighbourhood to implement a message propagation algorithm followed by a non linear activation function. It is defined as :
\begin{equation}
    Z^{(l)} = \delta(\hat{A}\cdot Z^{(l-1)},\theta^{(l)})
\end{equation}

Where a non-linearity function is denoted by $\delta$ (e.g ReLU), $\hat{A}$ represents normalised matrix A which is shown in the following:

\begin{equation}
    \hat{A} = D^{-\frac{1}{2}}(A+I)D^{\frac{1}{2}}
\end{equation}

where \begin{math} D_{ii} = \sum_{j} (A+I)_{ij} \end{math} is a diagonal matrix representing node degrees and I represents the identity matrix.

\subsubsection{Inference Module}
The inference module is mainly responsible for combining features from the CNN module and the GNN module. The skip connection from the CNN module are merged with the graph embeddings sequentially. Each concatenation operation is followed by a convolutional layer,activation layer, up-sampling operation and a  dropout layer acting as a regulariser. This process doubles the size of the tensors and maintains the channel size to 16 to facilitate smooth concatenation of embeddings.The dimensions of the the outputs of each block is made to match with the dimensions of the incoming skip connections from the CNN module. The penultimate convolutional layer reduces the channel size to 1 which results in the implicit combination of all features. The logits are then passed through a final sigmoid activation layer to produce 512X512 sized segmented maps.

\section{Results and Discussions}

The whole network was trained on a Nvidia P100 Graphical Processing Unit (GPU) for around 250 epochs for each experimental run. 318 CT scan slices were split into a training set containing 268 images and a test set containg 50 images. The model was built and trained using pytorch-lightning \cite{falcon2019pytorch}. Adam \cite{kingma2017adam} with a learning rate of \begin{math} 1 \times 10^{-4} \end{math} was used for finding a local minimum in the loss landscape. Dropout probability was set at 0.6 throughout the network. The model was optimised using different objective functions described in the following sections.

\subsection{Objective Functions}

In this section, we describe different objective functions commonly used for the task of semantic segmentation and their corresponding effectiveness when used in our study.

\subsubsection{Dice Loss}

Dice loss \cite{201712} is extensively used in medical segmentation literature as it helps redress the data imbalance problem in medical images. dice coefficient can be formally represented as :

\begin{equation}
    dice\:coefficient = \frac{2 \times TP}{2 \times TP + FP + FN}
\end{equation}

Where TP FP and FN are True Positive, False Positive and False Negative values respectively. Dice loss can be attributed as :

\begin{equation}
    dice\:loss = 1- dice\:coefficient
\end{equation}

\subsubsection{Binary Cross Entropy Loss}

Binary Cross Entropy (BCE) loss calculates a score based on the distance from an expected value that penalizes the probabilities based on their predicted probability. BCE loss can be formulated as follows:
\begin{equation}
    BCE\:loss = -\frac{1}{N}\sum_{i=1}^{N}y_{i}\cdot \log(\hat{y}) + (1-y_{i})\cdot \log(1-\hat{y})
\end{equation}

where  \begin{math} N,y,\hat{y}  \end{math} represent number of samples, ground truth labels and predictions respectively.

\subsubsection{Focal Tversky Loss}
 Tversky Index (TI) is a generalisation of dice loss. It adds weights to the FN (False Negatives) and FP (False Positive) terms for dice loss. TI  can be formulated as follows:

\begin{equation}
    TI(y,\hat{y}) = \frac{y \hat{y}}{y \hat{y} + (1-\beta)y(1-\hat{y}) + \beta \hat{y} (1-y)}
\end{equation}

Focal Tversky Loss (FTL) can be represented as follows :

\begin{equation}
    FTL = \sum (1-TI)^{\gamma} 
\end{equation}

where \begin{math}
\gamma
\end{math} varies in the range [1-3].

\subsubsection{Results For Loss Functions}
\begin{center}

\begin{tabularx}{0.8\textwidth} { 
  | >{\centering\arraybackslash}X 
  | >{\centering\arraybackslash}X 
  | >{\centering\arraybackslash}X | }
 \hline
 Objective Function & Dice Coefficient & Focal Tversky Loss \\
 \hline
 Dice Loss  & 0.75  & 0.37  \\
\hline
 BCE Loss  & 0.79  & 0.29  \\
 \hline
 Dice + BCE Loss & 0.81 & 0.35 \\
 \hline
 Focal Tversky Loss & 0.73 & 0.37 \\
 \hline

\end{tabularx}

\captionof{table}{Results for Different Objective Functions}

\label{tab: onjfunc}

\end{center}

We obtain a maximum dice score of 0.81 using a combination of Dice loss and BCE loss. Using Focal Tversky loss as an objective function did not provide a significant upgrade for our dice score with a reported dice score 0.73 the lowest we achieved in our study. BCE loss proved to be an effective objective function with a dice score of 0.79. Optimizing the model on dice loss alone did not prove to be an efficient solution as it provided a dice score of 0.75.

\subsection{Comparison With Different Models}

\begin{center}

\begin{tabularx}{0.8\textwidth} { 
  | >{\centering\arraybackslash}X 
  | >{\centering\arraybackslash}X 
  | >{\centering\arraybackslash}X|}
 \hline
 Model Citations & Data-set Samples & Dice Coefficient  \\
 \hline
 Hssayeni et al.  \cite{hssayeni2019intracranial}  & 254 & 0.677   \\
\hline
 Chang et al. \cite{pmid30049723} & 40000 & 0.85  \\
 \hline
 Kuang et al. \cite{kuang2019segmenting}  & 720  & 0.65 \\
 \hline
 Cho et al. \cite{cho2019affinity} & 6000 & 0.62 \\
 \hline
 Wang et al. \cite{app10093297} & 254 & 0.67 \\
 \hline
 Proposed  & 258 & 0.813 \\
 \hline

\end{tabularx}
\captionof{table}{Results for Different Models}
\label{tab: onjfunc}

\end{center}

Here we present a comparison for different models trained on varying amounts of samples and reported a dice score. Our model provides a dice coefficient of 0.81 using around half the number of parameters of a baseline U-Net architecture. Hasseyani et al. \cite{hssayeni2019intracranial} presented a baseline U-Net for a semantic segmentation task on the dataset \cite{6}. They achieved a dice coefficient of 0.67. Chang et al. trained on about 40000 collected samples. The proposed CNN based network reported a dice score of 0.85. Wang et al. proposed an attention-based semi-supervised U-Net based model and reported a dice score of 0.67. Kuang et al. proposed an ensemble of U-Net based architecture along with convex optimization-based techniques which resulted in a dice score of 0.65. Cho et al. propose an affinity graph-based architecture. They reported a dice score of 0.67.

\section{Conclusion}
\begin{figure}[h]

\includegraphics[width=8cm]{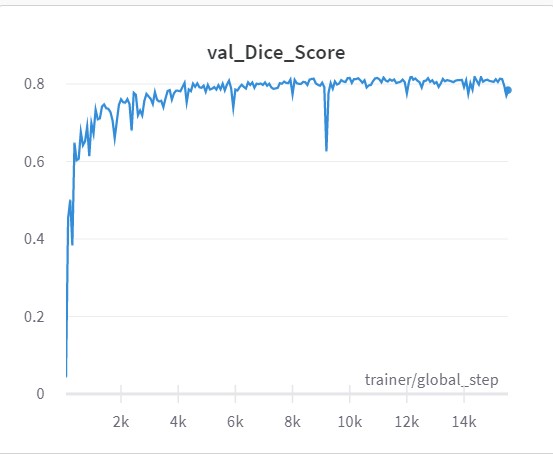}
\centering
\caption{Validation Dice Score V/s Training Steps}
\label{fig:val_dice}
\end{figure}

\begin{figure}[h]

\includegraphics[width=8cm]{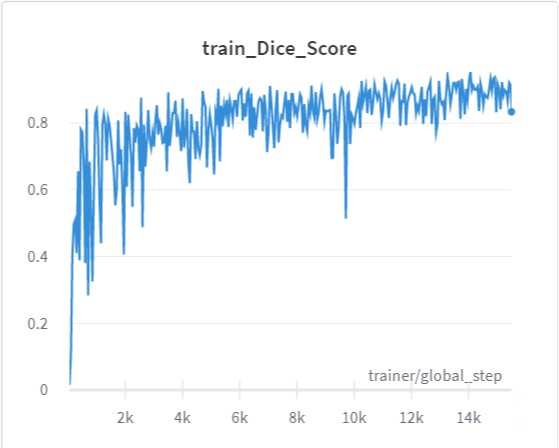}
\centering
\caption{Training Dice Score V/s Training Steps}
\label{fig:train_dice}
\end{figure}

In this study, we propose a novel GNN based architecture for the task of semantic segmentation of brain haemorrhages. We provide a description of three modules namely a CNN module a GNN module and an Inference module. We further provide a detailed description of a Self Constructing Graph (SCG) module. The SCG is used in conjunction with a Graph Convolutional Network Layer. The inference module is responsible for the combination of all information aggregated from the GNN and CNN modules. We further provide details about our training procedure and the different objective functions used for our study. We then proceed to conclude the study with a discussion of results achieved by different models.
A limitation this study faced was the lack of quality data. dataset. \cite{6} contains about 318 segmentation masks which are not enough data for the model. Our model suffers from mild overfitting as is evident from figures \ref{fig:val_dice} and \ref{fig:train_dice}. Our training dice score curve begins to diverge from the validation dice score curve which are signs of overfitting. Future works on study can include training and validating our architecture on more data which can provide an even more robust estimate for the performance of the model. Studies can also incorporate different GNN layers e.g. GraphSage \cite{hamilton2018inductive} and Graph Attention Layer (GAT) \cite{hi}.

\bibliographystyle{unsrt}  
\bibliography{references}

\end{document}